\documentstyle[11pt,paspconf,psfig]{article}

\begin{document}

\title{X-Ray Observations of GRS 1915+105}

\author{J. Greiner}
\affil{Max-Planck-Institute for extraterrestrial Physics, 85740 Garching, FRG}

\author{B.A. Harmon, W.S. Paciesas}
\affil{Marshall Space Flight Center, Huntsville, AL 35812, USA}

\author{E.H. Morgan, R.A. Remillard}
\affil{Center for Space Research, MIT, Cambridge, MA 02139, USA}


\section{Spectral analysis of simultaneous ROSAT/BATSE data}

\begin{figure}[hb]
     \vspace{-0.15cm}
     \centering{
     \hspace{0.01cm}
      \vbox{\psfig{figure=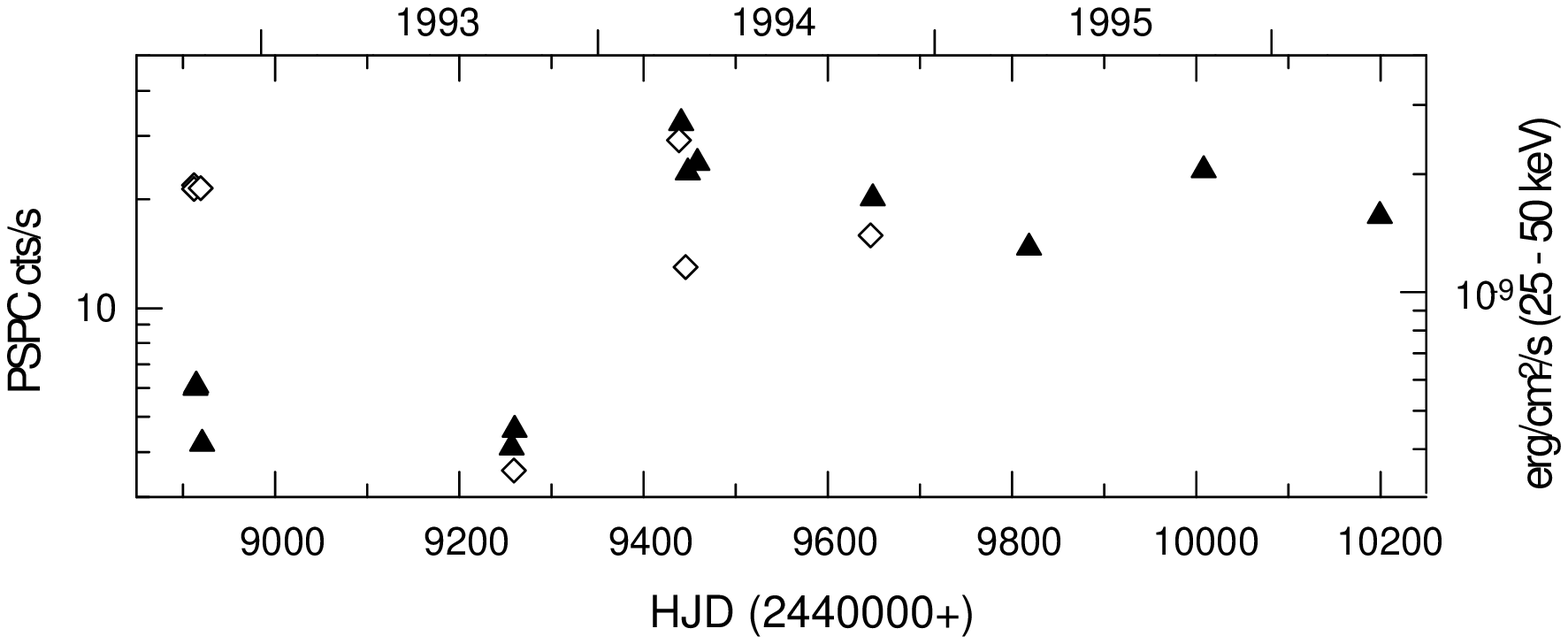,width=0.87\textwidth,%
       bbllx=1.9cm,bblly=3.2cm,bburx=20.cm,bbury=10.6cm,clip=}}\par}
     \vspace{-0.4cm}
    \caption[lc]{\small Intensities in the soft 0.1--2.4 keV ROSAT band (filled 
        triangles, left axis) and the 25--50 keV BATSE band (open squares,
        right axis). }
      \label{lc}
      \vspace{-0.3cm}
   \end{figure}

\begin{figure}[th]
      \vspace{-0.2cm}
     \centering{
     \hspace{0.01cm}
      \vbox{\psfig{figure=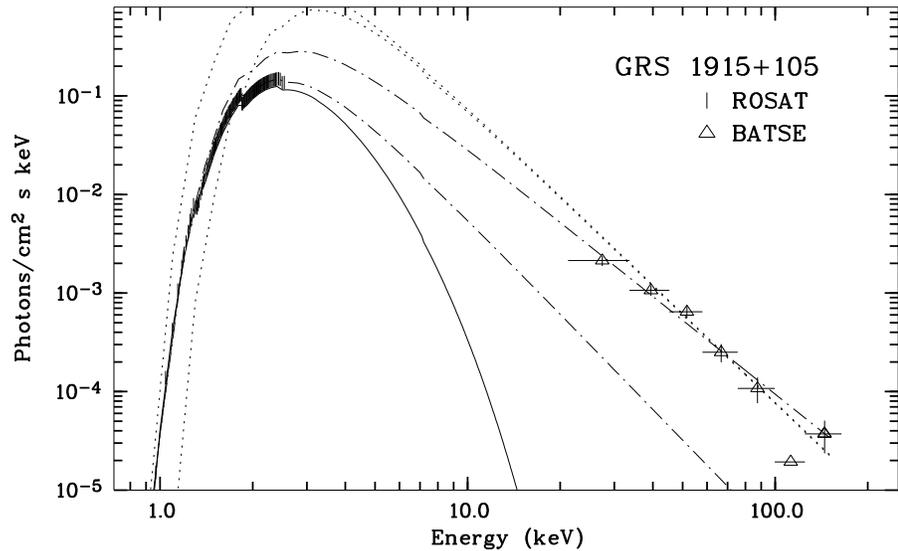,width=0.89\textwidth,%
       bbllx=1.9cm,bblly=16.9cm,bburx=18.4cm,bbury=27.2cm,clip=}}\par}
     \vspace{-0.5cm}
     \caption[fit]{\small Spectral fits of simultaneous ROSAT/BATSE data 
            of 19/20 October 1992 (see text). 
            This demonstrates that none of the single component models
            satisfactorily fits the combined simultaneous data.}
      \label{fit}
     \vspace{-0.3cm}
   \end{figure}

After the discovery of GRS 1915+105 (Castro-Tirado et al. 1992)
we obtained pointed ROSAT observations every six months (12 until now).
The flux in the ROSAT (0.1--2.4 keV) band is strikingly different from the
simultaneous BATSE (25--50 keV) flux which was obtained by integrating the
best fit power law (Fig. \ref{lc}).
Motivated by the different intensity evolution in the soft and hard
X-ray band we have selected BATSE monitoring data collected simultaneously
to ROSAT data and performed joint spectral fitting with XSPEC.
As a result, we never got an acceptable fit (see Fig. \ref{fit}):
The BATSE power law (upper dotted line) is too steep to match the ROSAT band,
and even allowing for an increased absorbing column (lower dotted line) 
does not solve the problem. Alternatively, neither a thermal bremsstrahlung 
fit (solid line) nor a power law fit (lower dash-dot line)
to the ROSAT data match the BATSE flux. The upper dash-dot line is a --2.5 
powerlaw which would match the BATSE data while giving too much 1--2 keV 
emission. A similar, but less stringent result is obtained when folding the
best fit BATSE power law models with the HRI detector response to compare the
expected count rate with the observed one.
We therefore conclude that the spectrum during all simultaneous ROSAT/BATSE
observations seemingly consists of two different spectral components.

\section{Pointed RXTE Observations}

Immediately after the discovery of the high activity level of GRS 1915+105
in the RXTE ASM data it was observed in a series of pointings starting
on April 6, 1996.
Details of the drastic X-ray intensity changes on a variety of time scales
ranging from sub-seconds to days are given elsewhere (Greiner, Morgan 
\& Remillard 1996). Summarizing, the observations over a time span of 13 weeks 
reveal 3 episodes (April 6, May 20--26, June 11--19)
of large amplitude variations between 0.2 and 3 Crab. 
In particular, the source exhibits quasi-periodic brightness sputters with
varying duration and repetition time scale. 
In between these episodes GRS 1915+105 remains at $\approx$1 Crab with  smaller
amplitude variations, but higher frequency.
The spectrum during the brightness sputters is remarkably different from
the spectrum of the mean high state emission.
We argue that such sputtering episodes are
possibly caused by a major accretion disk instability.
Based on the coincidence in time of two radio flares (Pooley 1996)
following the observed X-ray sputtering episodes
we speculate that superluminal ejections (as observed from GRS 1915+105
during earlier activity periods) are related to episodes of large
amplitude X-ray variations.

\acknowledgments
JG is supported by the German Bundesmi\-ni\-sterium f\"ur Bildung,
Wissenschaft, Forschung und Technologie under contract No.
FKZ 50 OR 9201. EHM and RR were supported by NASA contract NAS5-30612.

\end{document}